\documentstyle[12pt]{article}
\begin {document}
\newcommand {\beqa}{\begin{eqnarray}}
\newcommand {\eeqa}{\end{eqnarray}}
\newcommand {\n}{\nonumber \\}
\newcommand {\beq}{\begin{equation}}
\newcommand {\eeq}{\end{equation}}
\newcommand {\om}{\omega}

\newcommand {\de}{\delta}

\newcommand {\th}{\theta}

\newcommand {\Ga}{\Gamma}
%%%%%%%%%%%%%%%%%%%%%%%%%%%%%%%%%%%%%%%%%%%%%%%%%%%%%%%%%%%%%%%%%%%%%
\begin{flushright}
{\bf RI-10}
\\
November  1994
\vspace{1.0 in}\\
\end{flushright}
\begin{center}
{\large\bf Fundamental S-matrix\\
              for \\
      Vector Perturbed $WD_n$ Minimal Models} \vspace{.7 in}\\
{\bf A. Babichenko
        \vspace{.4 in}\\
        {\it Racah institute of Physics, Hebrew University }\\
        {\it Jerusalem, 91904, Israel  \vspace{.7 in}}}
\end{center}
\begin{abstract}
   Kink-kink S-matrix for integrable vector perturbed
$WD_{n}^{(k)}$ minimal
models is constructed from the Boltzmann weights of $A_{2n-1}^{(2)}$
 RSOS model and checked in two limit cases of $k$.
\end{abstract}
%%%%%%%%%%%%%%%%%%%%%%%%%%%%%%%%%%%%%%%%%%%%%%%%%%%%%%%%%%%%%%%%%%%%%%%
\newpage

{\bf 1.}In the last years,
starting from the famous work \cite{ZAM} where the main principles
of exact S-matrix construction for integrable perturbations of
Conformal Field Theories were formulated and demonstrated on few
examples, many other such integrable perturbations were found
and their S-matrices were studied and classified (see the review
\cite{MUS}). It seems that probably all
such massive
integrable perturbations of Virasoro minimal models (without
additional infinite symmetries) were found and S-matrices for them
were constructed,as far as the spectrum of such perturbations seems
 to be more rich and S-matrices much more complicated for the
perturbations of Conformal Field Theories with additional affine
symmetries. Most of such theories (if not all of them)
 may be expressed as coset constructions of some Kac-Moody algebras
at certain levels. Among this theories there are some integrable
perturbations of WZW models \cite{BONO}, $Z_n$
 parafermions \cite{FAT}, the adjoint perturbations
of W-invariant theories \cite{deVFAT} built on $A_n$ series of Lie
algebras (see, for example,\cite{FATLUK},\cite{SCHOUT}) and for
other series \cite{HOLLOW}.

Recently another class of integrable perturbations of W-invariant
theories were found \cite{VAY},\cite{YA}. In \cite{YA} it was checked
by "counting argument" \cite{ZAM}, that vector perturbation (in the
classification of primary fields according to \cite{FATLUK}) of
$WD_{n}^{(k)}$ minimal models are integrable and can be considered
as generalization of $(1,2)$  integrable perturbation of
Virasoro minimal models to the W-invariant theories. It was shown
there by explicit construction of nonlocal currents, that the model
has $A_{2n-1,q}^{(2)}$ quantum group of symmetry with
$q=e^{-\frac{i\pi}{2n-2+k}}$  and  therefore the same group of
S-matrix symmetry. The connection of the R-matrix commuting with the
constructed coproduct with the known R-matrix solution for this
affine group of symmetry \cite{JIM} was established and possible
ways of S-matrix construction on the base of this R-matrix were
discussed. In this note we suggest the fundamental S-matrix for
this model (kink-kink S-matrix) constructed on the base of Boltzmann
weights for vector representation of $A_{2n-1}^{(2)}$-invariant
 RSOS model and we check them for two particular cases of $k$ -
number of minimal model.

{\bf 2.}Before we present the conjectured S-matrix
recall the expressions for the central charge of $WD_{n}^{(k)}$
minimal models
($\sim SO(2n)_{k}\times SO(2n)_{1}/ SO(2n)_{k+1}$) and conformal
 dimension
of our perturbing primary field, corresponding to the fundamental
weight of vector representation of $D_n$

\beqa
c&=&n\left( 1-\frac{(2n-2)(2n-1)}{(2n-2+k)(2n-1+k)}\right)\label{c}\\
\Delta&=&\frac{1+(n-1+k)^{2}+\sum_{k=2}^{n-2} k^2}{2(2n-2+k)(2n-1+k)}
{}~~~~(n\geq 4) \label{dim} \\
\Delta&=&\frac{1+(2+k)^{2}}{2(4+k)(5+k)}
{}~~~~(n= 3) \nonumber
\eeqa\\
Let us point here two remarkable facts which will be used and
discussed later.  In the limit $k\rightarrow
\infty$ conformal dimension of the perturbation is going to 1/2.
Another feature is that for each $n$ the central charge of the lowest
minimal model $k=1$ is equal to 1.

   We will write down now the Boltzmann weights of  $A_{2n-1}^{(2)}$
 RSOS model based on the realization for this algebra made by using
the $D_n$ loop algebra \cite{WAR} (in contrast to $C_n$ realized
 $A_{2n-1}^{(2)}$ Boltzmann weights which was constructed in
\cite{KUN}). We will write them for the restricted model in the
trigonometric limit. Let us fix some notations. $\Lambda_i$ $(0\leq i
\leq n)$ denote the fundamental weights of $D_{n}^{(1)}$, and $\rho =
\Lambda_0+...+\Lambda_n$. Let ${\cal A}$ be the set of weights in
the vector representation of $D_n$ and for $a \in {\cal H}^*\equiv
\sum_{i=0}^{n} {\bf C} \Lambda_i$ we write $\bar{a}$ to mean its
classical  part. In terms of the orthogonal vectors $e_i$ $(1\leq
i\leq n)$, $(e_i,e_j)=\delta_ij$, $e_{-i}=-e_i$ classical parts
 $\bar{\Lambda}_i, \bar{\rho}$ and ${\cal A}$ can be written as
follows

\beqa
\bar{\Lambda}_i &=& e_1+ \cdots +e_i~~~~(1\leq i\leq n-2) \label{fv}\\
\bar{\Lambda}_{n-1} &=&\frac{1}{2}( e_1+ \cdots
+e_{n-1}-e_{n}) \n
\bar{\Lambda}_n &=&\frac{1}{2}( e_1+ \cdots
+e_{n-1}+e_{n}) \nonumber
\eeqa
\beqa
a&=&(L-a_1-a_2-1) \Lambda_0 +\sum_{i=1}^{n-1}(a_i-a_{i+1}-1)\Lambda_i
 +(a_{n-1}+a_n-1)\Lambda_n,  \label{heit}\\
L&>& a_1+a_2,~~a_1> a_2>\cdots > a_n,~~ a_{n-1}+a_n> 0 \nonumber
\eeqa\\
where $a_i\in {\bf Z}$ or $a_i\in {\bf Z}+\frac{1}{2}$ and $L=2n-2+k,
{}~~~(k=1,2,...)$ -- is the number of minimal unitary $WD$ model which we
perturbe. It can be easily  seen  that

\beq
\bar{a}+\bar{\rho} = \sum_{i=1}^{n} a_i e_i,~~~~a_{\mu}=< a+\rho,
e_{\mu}>, ~~~~-n\leq\mu\leq n. \nonumber
\eeq\\
It was shown in \cite{WAR} that Boltzmann weights (here we write them
in the trigonometric limit, while in \cite{WAR} they are written in
general elliptic form) for this RSOS model take the form

\beqa
[x]=sin \om x,~~~~[x]_{+}&=&cos\om x,~~~~
\om=\frac{\pi}{L} \nonumber\\
 W_{u} \left(\begin{array}{rr} a &a+e_{\mu}\\
a+e_{\mu} &a+2e_{\mu}
\end{array} \right)&=&\frac{[1+u][n+u]_{+}}{[1][n]_{+}} ~~~~~~
(\mu\neq 0) \label{bw}\\
 W_{u} \left(\begin{array}{rr} a &a+e_{\mu}\\ a+e_{\mu} &a+e_{\mu}+
e_{\nu}\end{array}
\right)&=&\frac{[a_{\mu\nu}-u][n+u]_{+}}{[a_{\mu\nu}][n]_{+}} ~~~~~~
(\mu\neq\pm\nu) \n
 W_{u} \left(\begin{array}{rr} a &a+e_{\nu}\\ a+e_{\mu} &a+e_{\mu}+
e_{\nu}\end{array}
\right)&=&\left(\frac{[a_{\mu\nu}+1][a_{\mu\nu}-1]}{[a_{\mu\nu}]^2}
\right)^{1/2}
\frac{[u][n+u]_{+}}{[1][n]_{+}} ~~~~~~
(\mu\neq\pm\nu) \n
 W_{u} \left(\begin{array}{rr} a &a+e_{\nu}\\ a+e_{\mu} &a\end{array}
\right)&=&\left(G_{a,\mu}G_{a,\nu}\right)^{1/2}
\frac{[u][a_{\mu-\nu}+1-n-u]_{+}}{[a_{\mu-\nu}+1][n]_{+}} ~~~~~~
(\mu\neq\nu) \nonumber
\eeqa

\beqa
 W_{u} \left(\begin{array}{rr} a &a+e_{\mu}\\ a+e_{\mu} &a\end{array}
\right)&=&\frac{[2a_{\mu}+1-u][n+u]_{+}}{[2a_{\mu}+1][n]_{+}} +
\frac{[u][2a_{\mu}+1-n-u]_{+}}{[2a_{\mu}+1][n]_{+}}G_{a,\mu} ~~
(\mu\neq 0) \n
&=&\frac{[2a_{\mu}+1-2n-u][n-u]_{+}}{[2a_{\mu}+1-2n][n]_{+}} -
\frac{[u][2a_{\mu}+1-n-u]_{+}}{[2a_{\mu}+1-2n][n]_{+}}H_{a,\mu}
 \nonumber
\eeqa\\
where $a_{\mu\nu}=a_{\mu}-a_{\nu}$, $a_{\mu-\nu}=a_{\mu}+a_{\nu}$,

\beqa
G_{a,\mu}&=&G_{a+e_{\mu}}/G_{a}=\left\{ \begin{array}{r} \prod_{k\neq
 0,\pm\mu} \frac{[a_{\mu k}+1]}{[a_{\mu k}]}~~~~~~\mu\neq 0\\ 1~~~~~~~~
{}~~~~~~\mu=0\end{array} \right. \label{GH}\\
G_{a}&=&\prod_{1\leq i<j\leq n} [a_i-a_j][a_i+a_j] \n
H_{a.\mu}&=&\sum_{k\neq\mu}\frac{[a_{\mu}+a_k +1-2n]}{[a_{\mu}+a_k +1]}
G_{a,k} \nonumber
\eeqa\\

Unitarity and crossing relations for these Boltzmann weights read as

\beqa
\sum_{g}W_u \left(\begin{array}{rr} a &g\\ c &d
\end{array}\right) W_{-u}\left(\begin{array}{rr} a &b\\ g &d
\end{array}\right)&=& \de_{bc}\frac{[n-u]_{+}[n+u]_{+}[1+u][1-u]}{
[n]_{+}^2 [1]^2}=\de_{bc}\rho (u) \label{un}\\
W_u \left(\begin{array}{rr} a &b\\ c &d
\end{array}\right)&=&\left(\frac{G_b G_c}{G_a G_d}\right)^{1/2} W_{
k/2-1-u} \left(\begin{array}{rr} c &a\\ d &b
\end{array}\right) \label{cr}
\eeqa\\

We would like now to construct the S-matrix for kinks on the base of
the solutions of Yang-Baxter equation written above. We denote a kink
state by $K_{ab}(\th)$, where $a$ and $b$ are two vacua of the theory
and $\th$ is the rapidity of the kink, and, according to the main
feature of two dimensional integrable field theory, we need only to
consider the S-matrix for the  process $K_{ac}(\th_1)+K_{cd}(\th_2)
\rightarrow K_{ab}(\th_2)+K_{bd}(\th_1)$, since all the other S-matrix
 elements are determined in terms of these. The idea is well known
\cite{deVFAT}\cite{HOLLOW}:
 we look for the S-matrix of the scattering process
of kinks in the form

\beq
S_{u}\left(\begin{array}{rr} a &b\\ c &d\end{array}\right)=
Y(u) W_{\eta u}\left(\begin{array}{rr}
a &b\\ c &d\end{array}\right)\left(\frac{G_a G_d}{G_b
G_c}\right)^{u/2} \label{sm}
\eeq\\
with  some scalar function $Y$ to be found, where $u$ is connected to
 the rapidity difference of the incoming kinks $\th$ by $u=\th/\pi i$,
and $\eta$ -- some constant.

The unitarity constraint can be satisfied by virtue of the relation
(\ref{un}) provided

\beq
Y(u)Y(-u)= 1/\rho(\eta u)  \label{uni}
\eeq\\
The crossing relation is satisfied provided $\eta$ is equal to the
crossing parameter $\eta=k/2-1$ and

\beq
Y(u)=Y(1-u) \label{cro}
\eeq\\
The nontrivial feature of the crossing parameter of the
Boltzmann weights for $A_{n}^{(2)}$ algebras based on the orthogonal
groups, is the presence of special point $k=2$, when it is equal
 to zero. This property objects to a construction of physical
scattering theory  directly  on the base of these Boltzmann
weights for the perturbed $k=2$ minimal model, although at the first
glance there is no something special in the $k=2$ case.
Having no an answer on this question, we meanwhile are going to
propose the fundamental S-matrix solution for all other $k$, supposing
 that $k=2$ case is the point of "regime change" for the S-matrix.

{\bf 3.} The system of functional equations (\ref{uni}) and (\ref{cro})
can be solved by standard iteration procedure, which has the
ambiguity of the first step giving rise to the well known CDD ambiguity
 of the solution. In what follows we will consider, without lost of
generality, the particular case $n=3$.

For the lowest minimal model $k=1$ we chose the first "test"
function for the iteration as a product of two gamma
functions divided by $cos$, which gives the following solution for $Y$:

\beqa
Y(u)=\frac{[1][3]_{+}\Ga( \frac{1}{5}(1-u/2)) \Ga ( 1-
\frac{1}{5}(1+u/2))}{[3+u/2]_{+}} \prod_{l=0}^{\infty}\frac{\Ga
( \frac{1}{5}(\frac{1}{2}-l+u/2))}{\Ga(
\frac{1}{5}(\frac{1}{2}-l-u/2))}
\frac{\Ga(\frac{1}{5}(-l-u/2)}{\Ga(\frac{1}{5}(-l+u/2)} \label{Y}\\
 \frac{\Ga(1- \frac{1}{5}(\frac{1}{2}-l+u/2))}{\Ga(1-
\frac{1}{5}(\frac{1}{2}-l-u/2))}\frac{\Ga(1- \frac{1}{5}
(2+l+u/2))}{\Ga(1-
\frac{1}{5}(2+l-u/2))}\frac{[l+\frac{7}{2}+u/2]_{+}
[l+4-u/2]_{+}}{[l+\frac{7}{2}-u/2]_{+}
[l+4+u/2]_{+}} \nonumber
\eeqa\\
This choice can be argued by the following check. One can see
that the S-matrix of vector
perturbed $WD_{n}^{(k)}$ minimal theories for $k=1$ should take
the form of the
(nonrestricted) Sine-Gordon  S-matrix at the special value of
its coupling constant, since the central charge for
this $k$ is equal to 1 for each $n$. Moreover, since the dimension
of perturbation for $k=1$ (\ref{dim}) is inverse even number, we
should expect to have SG S-matrix at reflectionless point.
 Indeed, in the case $k=1$
and, for example, $n=3$   the
restriction condition (\ref{heit}) leaves only four possibilities
for the choice of $(a_3,a_2,a_1)$: $(3,1,0)$; $(2,1,0)$;
$(\frac{5}{2},\frac{3}{2},\frac{1}{2})$ and
 $(\frac{5}{2},\frac{3}{2},-\frac{1}{2})$. The first set of $a_i$
corresponds to the highest weight of vector ($v$)
 representation of $D_3$, the second --
of  scalar ($0$), and fourth and third -- to the highest weights of
two spinor representations ($s$ and $c$). Using the admissibility
condition, we have
the following four nonzero Boltzmann weights in this case

\beqa
 W_{u} \left(\begin{array}{rr} v &0\\ 0 &v\end{array} \right),
{}~~ W_{u} \left(\begin{array}{rr} 0 &v\\ v &0\end{array} \right),
{}~~ W_{u} \left(\begin{array}{rr} s &c\\ c &s\end{array} \right),
{}~~ W_{u} \left(\begin{array}{rr} c &s\\ s &c\end{array} \right)
\label{sgbv}
\eeqa\\
All of them are representatives of the last type of non zero Boltzmann
weights in (\ref{bw}) and explicit calculation gives the same
expression for all of them

\beqa
W(u)&=&\frac{[3-u]_{+}[1+u]}{[1][3]_{+}} \nonumber
\eeqa\\
and for this case the crossing factor $(G_a G_d/G_b G_c)^{u/2}$
turns out to be equal to one for all of four types of S-matrix.
Such a trivial  tensor structure is compatible with the tensor
structure of soliton-antisoliton  SG S-matrix \cite{ZAMZAM}

\beqa
S(\th,\xi)=S_{0}(\th,\xi)R(\th,\xi)=S_{0}(\th,\xi)
\left(\begin{array}{rrrr} sh(\frac{\pi}{\xi}(\th-i\pi))
&~~~~&~~~~&~~~~~\\
{}~~~~~ &-sh(\frac{i\pi^2}{\xi})&-sh(\frac{\pi}{\xi}\th)&~~~~~\\
{}~~~~~ &-sh(\frac{\pi}{\xi}\th)&-sh(\frac{i\pi^2}{\xi})&~~~~~\\
{}~~~~~&~~~~&~~~~&sh(\frac{\pi}{\xi}(\th-i\pi))
\end{array} \right) \label{sg}\\
=\left(\begin{array}{rrrr} S(\th,\xi)&~~~&~~~&~~~\\
{}~~~ &S_{R}(\th,\xi)&S_{T}(\th,\xi)&~~~\\
{}~~~ &S_{T}(\th,\xi)&S_{R}(\th,\xi)&~~~\\
{}~~~&~~~&~~~&S(\th,\xi)\end{array} \right)  \nonumber
\eeqa
\beqa
S_{0}(\th,\xi)&=&\frac{1}{sh(\frac{\pi}{\xi}(\th-i\pi))} exp
\left[ -i\int_0^{\infty} \frac{dx}{x}\frac{sin(x\th)sh(
\frac{\pi-\xi}{2}x)}{ch(\frac{\pi x}{2}) sh(\frac{\xi x}{2})}
\right] \label{S0}
\eeqa\\
at $\xi=\pi/5$ reflectionless point.
 Moreover, full  equivalence of these two S-matrices, including the
infinite product of $Y$ and exponential of integral of (\ref{S0}),
also can be established. (After we rewrite the cosines in the infinite
product (\ref{Y}) as the product of two gamma functions,
 partial cancelation of
gamma functions takes place and we leave with the infinite CDD type
product of 8 gamma functions, which, after some algebra,
 together with other factors, gives the $S_0$ of (\ref{S0})).
\\

   In the case $k>2$ solution for $Y$ should be taken in the
following form (as before we write the answer for $n=3$ case)

\beqa
Y(u)&=&\frac{[1][3]_{+}}{[1-\eta u][3-\eta u]_{+}} \prod_{l=0}^{\infty}
\frac{[2(l+1)-k(l+\frac{1}{2})-\eta u][2l+3-k(l+1)+\eta u]}
{[2(l+1)-k(l+\frac{1}{2})+\eta u][2l+3-k(l+1)-\eta u]} \label{Y3}\\
& &\frac{[2l+4-k(l+\frac{1}{2})-\eta u]_{+}[2l+5-k(l+1)+\eta u]_{+}}
{[2l+4-k(l+\frac{1}{2})+\eta u]_{+}[2l+5-k(l+1)-\eta u]_{+}}\nonumber
\eeqa\\
This choice is dictated by the following argument. Since in the
limit $k\rightarrow \infty$ the model  under consideration takes
the form of free fermions--$SO(2n)_1$ Kac-Moody perturbed by the field
of conformal dimension $1/2$ (see (\ref{dim})), we expect the trivial
limit (-1) for the S-matrix. The proposed solution for S-matrix seems
to be the only one with this property. Now we will describe this
limit of S-matrix.

  First of all, as it was pointed out in \cite{deVFAT},
for the parameters
$a_{\mu}$ this limit means that $a_{\mu}, a_{\mu\nu}\rightarrow\infty$,
 $a_{\mu}/k, a_{\mu\nu}/k\rightarrow 0$. One can show that the
infinite product in $Y$ goes to 1 in the limit $k\rightarrow\infty$.
It can easily be checked that the prefactor before the infinite
product in $Y$ together with the five types of Boltzmann weights
(\ref{bw})
gives the zero limit for all of them except for the first and
the third one, for which the limit is equal to -1, giving the -1
 limit for the S-matrix.

{\bf 4.} We expect the mass spectrum, particle content
(higher kinks and breathers)
 and structure of full S-matrix to be rather complicated
for general case. (Even the lowest model of general case
($n=3,k=3$) has 24 fundamental particles.) In the same
way as it was shown in \cite{KUN} one can show, that
the problem of spectral decomposition for
our R-matrix is equivalent to this problem for
two arbitrary representations obtained by tensor product
of vector representation of $D_n$ algebra, which is unknown
in general case. Some examples of bootstrap for $SO(n)$
symmetric R-matrix in the simplest particular cases
have been shown in \cite{MCKAY} and led to a complicated
picture.

Another interesting question is the understanding of
$k=2$ case and construction of the S-matrix for it.
It was checked by \cite{KUNWAR} that the naive
regularization of zero crossing parameter leads to
a trivial S-matrix.

So needless to say that there are many points in the
S-matrix construction which should be understood.

%%%%%%%%%%%%%%%%%%%%%%%%%%%%%%%%%%%%

\vskip 2.0cm
I  thank S.Elitzur, I.Vaysburd, A.Kuniba and O.Warnaar for helpful
discussions.


\begin{thebibliography}{999}

\bibitem{ZAM} A.B.Zamolodchikov  {\em  Advanced Studies in
Pure Mathematics} {\bf 19 (1989) 641};\\
V.A.Fateev and A.B.Zamolodchikov  {\em Int. Journal of Mod. Phys.}
{\bf A5 (1990) 1025}
\bibitem{MUS} G.Mussardo  {\em  Physics Reports} {\bf 218 (1993) 215}
\bibitem{BONO} L.Bonora, Y.Z.Zhang and M.Martellini
{\em Int. Journal of Mod. Phys.} {\bf A6 (1991) 1617}
\bibitem{FAT}  V.A.Fateev {\em Int. Journal of Mod. Phys.}
{\bf A6 (1991) 2109}
\bibitem{deVFAT} H.J. de Vega, V.A.Fateev
{\em Int. Journal of Mod. Phys.} {\bf A6 (1991) 3221}
\bibitem{FATLUK} V.A.Fateev, S.L.Lukyanov
 {\em Soviet Scientific Reviews}
{\bf A Phys. 15 (1990) 1}
\bibitem{SCHOUT}  P.Bouwknegt and K.Schoutens {\em Physical Reports}
{\bf 223 (1993) 183}
\bibitem{HOLLOW} T.J.Hollowood {\em Nucl.Phys}
{\bf B414 (1994) 379}
 \bibitem{VAY} I.Vaysburd {\em Physics Letters}
{\bf 335B (1994) 161}
\bibitem{YA} A.Babichenko
{\em Integrable vector perturbations of W-invariant theories\\
 and their quantum group symmetry}; Preprint RI-7, 1994.(hepth 9406197)
\bibitem{JIM}  M.Jimbo {\em Com.Math.Phys.}
{\bf 102 (1986) 537}
\bibitem{WAR} S.O.Warnaar
{\em Algebraic construction of higher rank dilute A models};\\
Preprint of Melbourne Univ., 1994
\bibitem{KUN}  A.Kuniba  {\em Nucl.Phys.}
{\bf B355 (1991) 801}
\bibitem{ZAMZAM}  A.B.Zamolodchikov and Al.B.Zamolodchikov
 {\em Ann. Phys.}{\bf 120 (1979) 253};\\
\bibitem{MCKAY} N.J.MacKay {\em Nuclear Physics}
{\bf 356B (1991) 729}
\bibitem{KUNWAR} A.Kuniba and S.Ole Warnaar
  Private communication.
\
\

\end{thebibliography}
\end{document}